\documentclass[aps,prd,twocolumn,showpacs,superscriptaddress,groupedaddress]{revtex4}  
\usepackage{graphicx}
\usepackage{graphics}
\usepackage{fancyhdr}
\usepackage{amsmath} 
\usepackage{bm}
\usepackage{amsxtra}
\usepackage{amssymb}
\usepackage{amsthm}
\usepackage{latexsym}
\usepackage{lscape}
\usepackage{subfigure}
\usepackage{setspace}

\usepackage{enumerate}

\begin{document}

\noindent%
\mbox{FERMILAB-PUB-15-080-E \hspace{15mm}{\em Published in Phys. Rev. D as DOI: 10.1103/PhysRevD.91.072008}}
\noindent%
\hspace{5.7in} \mbox{(accepted)}

\title{Measurement of the forward-backward  asymmetry in   {\boldmath $\Lambda_b^0$} and {\boldmath $\overline \Lambda_b^0$} baryon production in {\boldmath  $ p \overline p$} collisions at  {\boldmath $\sqrt s =1.96$}~TeV }

\affiliation{LAFEX, Centro Brasileiro de Pesquisas F\'{i}sicas, Rio de Janeiro, Brazil}
\affiliation{Universidade do Estado do Rio de Janeiro, Rio de Janeiro, Brazil}
\affiliation{Universidade Federal do ABC, Santo Andr\'e, Brazil}
\affiliation{University of Science and Technology of China, Hefei, People's Republic of China}
\affiliation{Universidad de los Andes, Bogot\'a, Colombia}
\affiliation{Charles University, Faculty of Mathematics and Physics, Center for Particle Physics, Prague, Czech Republic}
\affiliation{Czech Technical University in Prague, Prague, Czech Republic}
\affiliation{Institute of Physics, Academy of Sciences of the Czech Republic, Prague, Czech Republic}
\affiliation{Universidad San Francisco de Quito, Quito, Ecuador}
\affiliation{LPC, Universit\'e Blaise Pascal, CNRS/IN2P3, Clermont, France}
\affiliation{LPSC, Universit\'e Joseph Fourier Grenoble 1, CNRS/IN2P3, Institut National Polytechnique de Grenoble, Grenoble, France}
\affiliation{CPPM, Aix-Marseille Universit\'e, CNRS/IN2P3, Marseille, France}
\affiliation{LAL, Universit\'e Paris-Sud, CNRS/IN2P3, Orsay, France}
\affiliation{LPNHE, Universit\'es Paris VI and VII, CNRS/IN2P3, Paris, France}
\affiliation{CEA, Irfu, SPP, Saclay, France}
\affiliation{IPHC, Universit\'e de Strasbourg, CNRS/IN2P3, Strasbourg, France}
\affiliation{IPNL, Universit\'e Lyon 1, CNRS/IN2P3, Villeurbanne, France and Universit\'e de Lyon, Lyon, France}
\affiliation{III. Physikalisches Institut A, RWTH Aachen University, Aachen, Germany}
\affiliation{Physikalisches Institut, Universit\"at Freiburg, Freiburg, Germany}
\affiliation{II. Physikalisches Institut, Georg-August-Universit\"at G\"ottingen, G\"ottingen, Germany}
\affiliation{Institut f\"ur Physik, Universit\"at Mainz, Mainz, Germany}
\affiliation{Ludwig-Maximilians-Universit\"at M\"unchen, M\"unchen, Germany}
\affiliation{Panjab University, Chandigarh, India}
\affiliation{Delhi University, Delhi, India}
\affiliation{Tata Institute of Fundamental Research, Mumbai, India}
\affiliation{University College Dublin, Dublin, Ireland}
\affiliation{Korea Detector Laboratory, Korea University, Seoul, Korea}
\affiliation{CINVESTAV, Mexico City, Mexico}
\affiliation{Nikhef, Science Park, Amsterdam, the Netherlands}
\affiliation{Radboud University Nijmegen, Nijmegen, the Netherlands}
\affiliation{Joint Institute for Nuclear Research, Dubna, Russia}
\affiliation{Institute for Theoretical and Experimental Physics, Moscow, Russia}
\affiliation{Moscow State University, Moscow, Russia}
\affiliation{Institute for High Energy Physics, Protvino, Russia}
\affiliation{Petersburg Nuclear Physics Institute, St. Petersburg, Russia}
\affiliation{Instituci\'{o} Catalana de Recerca i Estudis Avan\c{c}ats (ICREA) and Institut de F\'{i}sica d'Altes Energies (IFAE), Barcelona, Spain}
\affiliation{Uppsala University, Uppsala, Sweden}
\affiliation{Taras Shevchenko National University of Kyiv, Kiev, Ukraine}
\affiliation{Lancaster University, Lancaster LA1 4YB, United Kingdom}
\affiliation{Imperial College London, London SW7 2AZ, United Kingdom}
\affiliation{The University of Manchester, Manchester M13 9PL, United Kingdom}
\affiliation{University of Arizona, Tucson, Arizona 85721, USA}
\affiliation{University of California Riverside, Riverside, California 92521, USA}
\affiliation{Florida State University, Tallahassee, Florida 32306, USA}
\affiliation{Fermi National Accelerator Laboratory, Batavia, Illinois 60510, USA}
\affiliation{University of Illinois at Chicago, Chicago, Illinois 60607, USA}
\affiliation{Northern Illinois University, DeKalb, Illinois 60115, USA}
\affiliation{Northwestern University, Evanston, Illinois 60208, USA}
\affiliation{Indiana University, Bloomington, Indiana 47405, USA}
\affiliation{Purdue University Calumet, Hammond, Indiana 46323, USA}
\affiliation{University of Notre Dame, Notre Dame, Indiana 46556, USA}
\affiliation{Iowa State University, Ames, Iowa 50011, USA}
\affiliation{University of Kansas, Lawrence, Kansas 66045, USA}
\affiliation{Louisiana Tech University, Ruston, Louisiana 71272, USA}
\affiliation{Northeastern University, Boston, Massachusetts 02115, USA}
\affiliation{University of Michigan, Ann Arbor, Michigan 48109, USA}
\affiliation{Michigan State University, East Lansing, Michigan 48824, USA}
\affiliation{University of Mississippi, University, Mississippi 38677, USA}
\affiliation{University of Nebraska, Lincoln, Nebraska 68588, USA}
\affiliation{Rutgers University, Piscataway, New Jersey 08855, USA}
\affiliation{Princeton University, Princeton, New Jersey 08544, USA}
\affiliation{State University of New York, Buffalo, New York 14260, USA}
\affiliation{University of Rochester, Rochester, New York 14627, USA}
\affiliation{State University of New York, Stony Brook, New York 11794, USA}
\affiliation{Brookhaven National Laboratory, Upton, New York 11973, USA}
\affiliation{Langston University, Langston, Oklahoma 73050, USA}
\affiliation{University of Oklahoma, Norman, Oklahoma 73019, USA}
\affiliation{Oklahoma State University, Stillwater, Oklahoma 74078, USA}
\affiliation{Brown University, Providence, Rhode Island 02912, USA}
\affiliation{University of Texas, Arlington, Texas 76019, USA}
\affiliation{Southern Methodist University, Dallas, Texas 75275, USA}
\affiliation{Rice University, Houston, Texas 77005, USA}
\affiliation{University of Virginia, Charlottesville, Virginia 22904, USA}
\affiliation{University of Washington, Seattle, Washington 98195, USA}
\author{V.M.~Abazov} \affiliation{Joint Institute for Nuclear Research, Dubna, Russia}
\author{B.~Abbott} \affiliation{University of Oklahoma, Norman, Oklahoma 73019, USA}
\author{B.S.~Acharya} \affiliation{Tata Institute of Fundamental Research, Mumbai, India}
\author{M.~Adams} \affiliation{University of Illinois at Chicago, Chicago, Illinois 60607, USA}
\author{T.~Adams} \affiliation{Florida State University, Tallahassee, Florida 32306, USA}
\author{J.P.~Agnew} \affiliation{The University of Manchester, Manchester M13 9PL, United Kingdom}
\author{G.D.~Alexeev} \affiliation{Joint Institute for Nuclear Research, Dubna, Russia}
\author{G.~Alkhazov} \affiliation{Petersburg Nuclear Physics Institute, St. Petersburg, Russia}
\author{A.~Alton$^{a}$} \affiliation{University of Michigan, Ann Arbor, Michigan 48109, USA}
\author{A.~Askew} \affiliation{Florida State University, Tallahassee, Florida 32306, USA}
% ...
%
%
\collaboration{The D0 Collaboration\footnote{with visitors from
$^{a}$Augustana College, Sioux Falls, SD, USA,
$^{b}$The University of Liverpool, Liverpool, UK,
$^{c}$DESY, Hamburg, Germany,
$^{d}$CONACyT, Mexico City, Mexico,
$^{e}$SLAC, Menlo Park, CA, USA,
$^{f}$University College London, London, UK,
$^{g}$Centro de Investigacion en Computacion - IPN, Mexico City, Mexico,
$^{h}$Universidade Estadual Paulista, S\~ao Paulo, Brazil,
$^{i}$Karlsruher Institut f\"ur Technologie (KIT) - Steinbuch Centre for Computing (SCC),
D-76128 Karlsruhe, Germany,
$^{j}$Office of Science, U.S. Department of Energy, Washington, D.C. 20585, USA,
$^{k}$American Association for the Advancement of Science, Washington, D.C. 20005, USA,
$^{l}$Kiev Institute for Nuclear Research, Kiev, Ukraine,
$^{m}$University of Maryland, College Park, Maryland 20742, USA
and
$^{n}$European Orgnaization for Nuclear Research (CERN), Geneva, Switzerland
}} \noaffiliation
\vskip 0.25cm
\date{April 27, 2015}

\begin{abstract}
We  measure the forward-backward asymmetry in the production of
$\Lambda_b^0$ and $\overline \Lambda_b^0$ baryons
as a function of rapidity in $p \overline p $ collisions at $\sqrt s =1.96$~TeV using $10.4$~fb$^{-1}$ of data collected with the D0 detector at the Fermilab Tevatron collider. The asymmetry is determined by  the preference of $\Lambda_b^0$ or  $\overline \Lambda_b^0$ particles to be produced 
in the direction of the beam protons or antiprotons, respectively. 
The measured asymmetry integrated over rapidity $y$ in the range $0.1<|y|<2.0$ is $A=0.04 \pm 0.07 {\rm \thinspace (stat)} \pm 0.02 {\rm \thinspace (syst)}$.
\end{abstract}


\maketitle

Hadroproduction of particles carrying a heavy quark $Q$ ($Q=b,c$)  proceeds
through  gluon-gluon fusion or  quark-antiquark annihilations~\cite{nde}, followed by  hadronization of the heavy quarks. At the parton level of
leading-order (LO) quantum chromodynamics (QCD), $Q$ and $\overline Q$ quarks are produced symmetrically. Next-to-leading order (NLO) QCD effects can  introduce a small
 asymmetry of $\approx1$\% in $Q$ and $\overline  Q$ momenta from interfering amplitudes. The hadronization process 
may also change the direction of the particle carrying the quark $Q$  relative to the original $Q$ direction and thus generate a significant asymmetry.

 There have been few studies of this effect in bottom baryon production compared to bottom mesons.  Production of heavy baryons is sensitive to effects of nonperturbative final state interactions of a QCD string connecting the $b$ quark and a remnant of the proton. The production of the ground-state bottom  baryon $\Lambda_b^0$ and its antiparticle $\overline \Lambda_b^0$ has been recently discussed by Rosner~\cite{rosner}, who proposes the ``string drag'' mechanism that may  
 favor production of $\Lambda_b^0$ baryons in the hemisphere containing the beam proton, and $\overline \Lambda_b^0$ baryons in the antiproton beam hemisphere.
In the string drag picture, the QCD interaction  between a $b$ quark produced in the $p \overline p$ collision and  the remnant of the proton is described by a string
 with a linear potential.  When the  string breaks, it imparts an impulse to the quark along the beam axis. 
 Assuming a  string tension of 0.18~GeV$^2$,  
Rosner made an approximate prediction for the shift in the particle longitudinal momentum relative to the axis along the beam direction of $\Delta p_z=1.4$~GeV, resulting in a
shift in rapidity of approximately $\Delta y =1.4\text{~GeV}/E$,  where $E$ is the energy of the particle and the rapidity is defined as $y=\ln((E+p_z)/(E-p_z))/2$.
Another possible source of  
asymmetry in  $\Lambda_b^0$ production is the coalescence of an intrinsic $b$ quark at large momentum fraction $x$ in the Fock state  $|uudb\overline b\rangle$  of the proton  with a comoving diquark $ud$  from the  proton~\cite{brodsky}. 

In this article, we present a study of the  forward-backward production asymmetry of $\Lambda_b^0$ and $\overline \Lambda_b^0$ baryons  using  the fully reconstructed decay chain  $\Lambda_b^0 \rightarrow J/\psi \Lambda$, $J/\psi \rightarrow \mu^+ \mu^-$, $\Lambda \rightarrow p \pi^-$, and its charge conjugate. 
The forward ($F$) category corresponds to a particle ($\Lambda_b^0$ or $\overline \Lambda_b^0$) sharing valence quark flavors with a beam particle with the same sign of rapidity, and the backward ($B$) category corresponds to the reverse association.
In $p \overline p$ collisions at D0, we choose the positive $z$-axis to be in the direction of the proton beam, so that the forward direction corresponds to a  $\Lambda_b^0$ particle emitted  with $y>0$ or a $\overline \Lambda_b^0$ particle emitted at $y<0$.
In $ p p$ collisions,  $\Lambda_b^0$ particles  are assigned to  the forward category and  $\overline \Lambda_b^0$ particles to the backward category.
To facilitate a comparison with existing  measurements, we present  the ratio
of the backward to forward production cross sections, $R=\sigma(B)/\sigma(F)$, and the forward-backward asymmetry,
$A=(\sigma(F)-\sigma(B))/(\sigma(F)+\sigma(B))$, as functions of the rapidity $y$.
The data sample corresponds to an integrated luminosity of 10.4~fb$^{-1}$ collected with the D0 detector  in $p \overline p $ collisions at $\sqrt{s} = 1.96$~TeV at the Fermilab Tevatron collider.

Using the same data set, the D0 experiment has studied the forward-backward asymmetry in the production of $B^{\pm}$ mesons,
observing  no rapidity dependence~\cite{d0bplus}. 
The measured forward-backward asymmetry in the production of $B^{\pm}$ mesons, 
where the forward category corresponds to   $B^-$ mesons
produced at $y>0$
and  $B^+$ mesons produced at $y<0$,  is
$A_{\rm FB}(B^{\pm}) =[-0.24  \pm 0.41 {\rm \thinspace (stat)} \pm 0.19 {\rm \thinspace (syst)}]$\%, integrated over rapidity.


The D0 detector consists of a central tracking system, calorimeters,  and
muon detectors~\cite{Abazov2006463}. The central
tracking system comprises  a silicon microstrip tracker and a central
fiber tracker, both located inside a 1.9~T superconducting solenoidal
magnet.  The tracking system is designed to optimize tracking and vertexing
for pseudorapidities $|\eta|<3$,
where  $\eta = -\ln[\tan(\theta/2)]$, and  $\theta$ is the 
polar angle with respect to the proton beam direction.
  The tracking system can reconstruct the primary $p\overline{p}$ interaction vertex 
for interactions   with at least three secondary tracks with a precision
of $\approx 35$~$\mu$m ($\approx 90$~$\mu$m) in the plane transverse to (along) the beam direction.
  The muon detector, positioned outside the calorimeter, consists of a central muon system covering the pseudorapidity region of $|\eta|<1$ and a forward muon system covering the pseudorapidity region of $1<|\eta|<2$. Both central and forward systems consist of a layer of drift  tubes
and scintillators inside 1.8~T iron toroidal magnets and two similar layers outside the toroids~\cite{d0mu}. 
The toroid and solenoid magnet polarities were periodically reversed, allowing for a cancellation of first-order effects related to a possible instrumental asymmetry.


Candidate events are required to include a pair of  oppositely charged muons.
 At least one  muon  is required to be detected
 in the muon chambers in front of and behind a toroid magnet.
The other muon may be detected only in front of the toroid or as a 
minimum ionizing particle  
in the calorimeter.
Each muon candidate is required to match a track found in the central tracking system.

To form $\Lambda_b^0$ and  $\overline{\Lambda}_b^0$ candidates, muon pairs in the invariant mass range $2.9<M(\mu^+ \mu^-)<3.3$~GeV, consistent with a $J/\psi$ meson decay, are combined with $\Lambda$ baryon candidates. 
The $\Lambda$ candidates are formed from pairs of oppositely
charged
tracks originating from a common vertex,  consistent with a decay $\Lambda \rightarrow p \pi^-$ or $\overline \Lambda \rightarrow \overline p \pi^+$.
The charged particle with the higher momentum is assigned the proton mass.
A previous analysis has shown that the misassignment of the proton track is negligible~\cite{Lblifetime}.
 The $\Lambda$ candidate is required to have an invariant mass between $1.107$ and $1.125$~GeV and a transverse momentum greater than $1.8$~GeV.
The  separation of the $\Lambda$ decay vertex from the primary vertex 
in the transverse plane must be between $0.5$ and $25$~cm. 
A kinematic fit of the parameters of tracks forming the  $\Lambda_b^0$ candidate is performed by constraining the dimuon invariant mass  to the world-average $J/\psi$ mass~\cite{pdg2014}, and constraining  the $J/\psi \Lambda$ system to originate from a common decay vertex.
The modified track parameters are used in the
calculation of  the $\Lambda_b^0$  invariant mass. 
We require  $5.0 <M(J/\psi \Lambda)<6.2$~GeV.

To suppress the large background from prompt $J/\psi$ production, we require a significant separation
of the $\Lambda_b^0$ decay vertex from the primary vertex.
To reconstruct the primary vertex, tracks are selected that do not 
belong to
the 
$\Lambda_b^0$ decay. 
We constrain the transverse position of the primary vertex to the average beam location in the transverse plane.
We define the signed  decay length of a $\Lambda_b^0$ baryon, $L_{xy}$, 
as the  vector pointing
from the primary vertex to the $\Lambda_b^0$ decay vertex projected  on the
direction of the $\Lambda_b^0$ transverse momentum $\vec{p}_T$.
We require $L_{xy}$ to be greater than three times its uncertainty.

\begin{figure}
  \centering
  {\includegraphics[width=0.9\columnwidth]{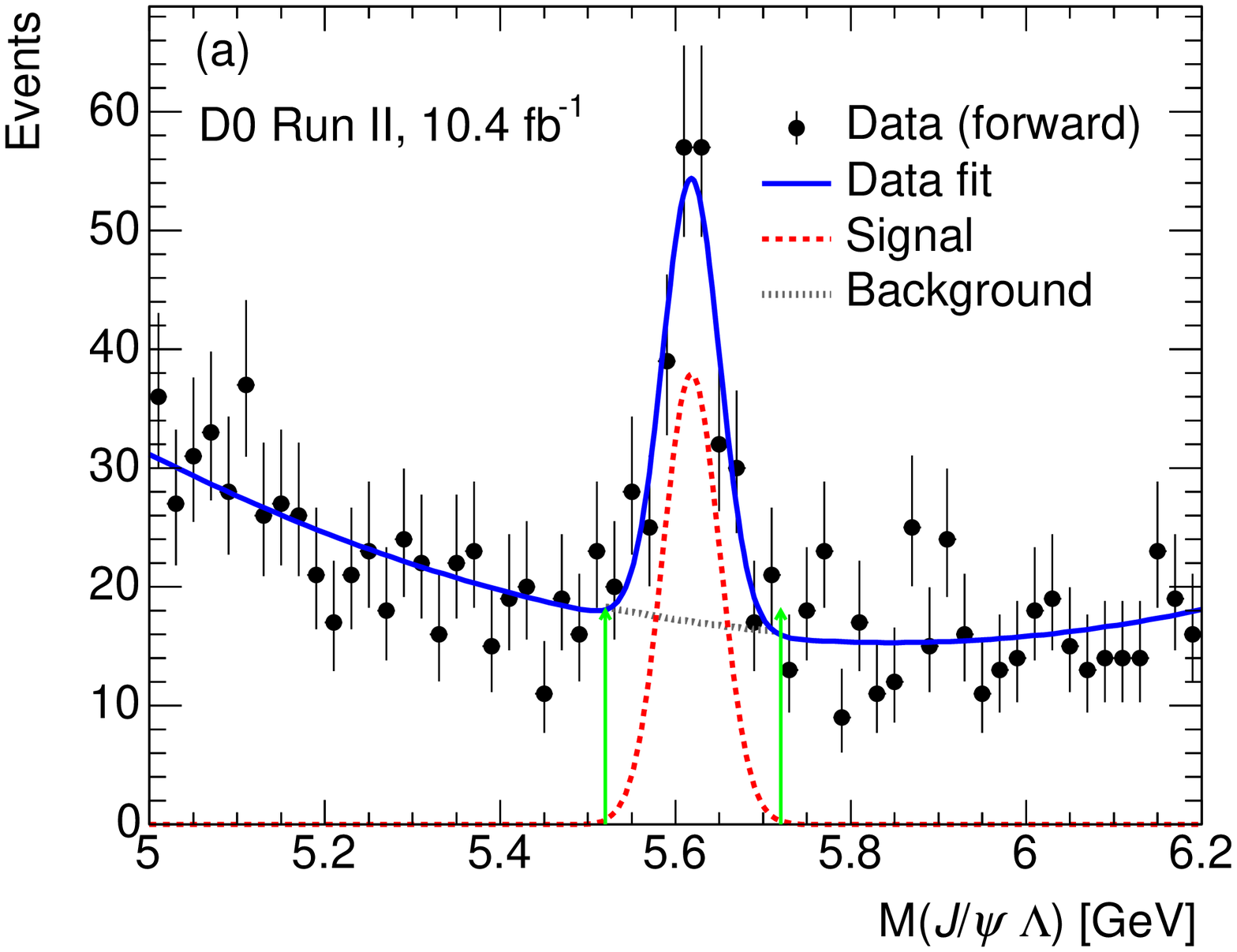}}
 {\includegraphics[width=0.9\columnwidth]{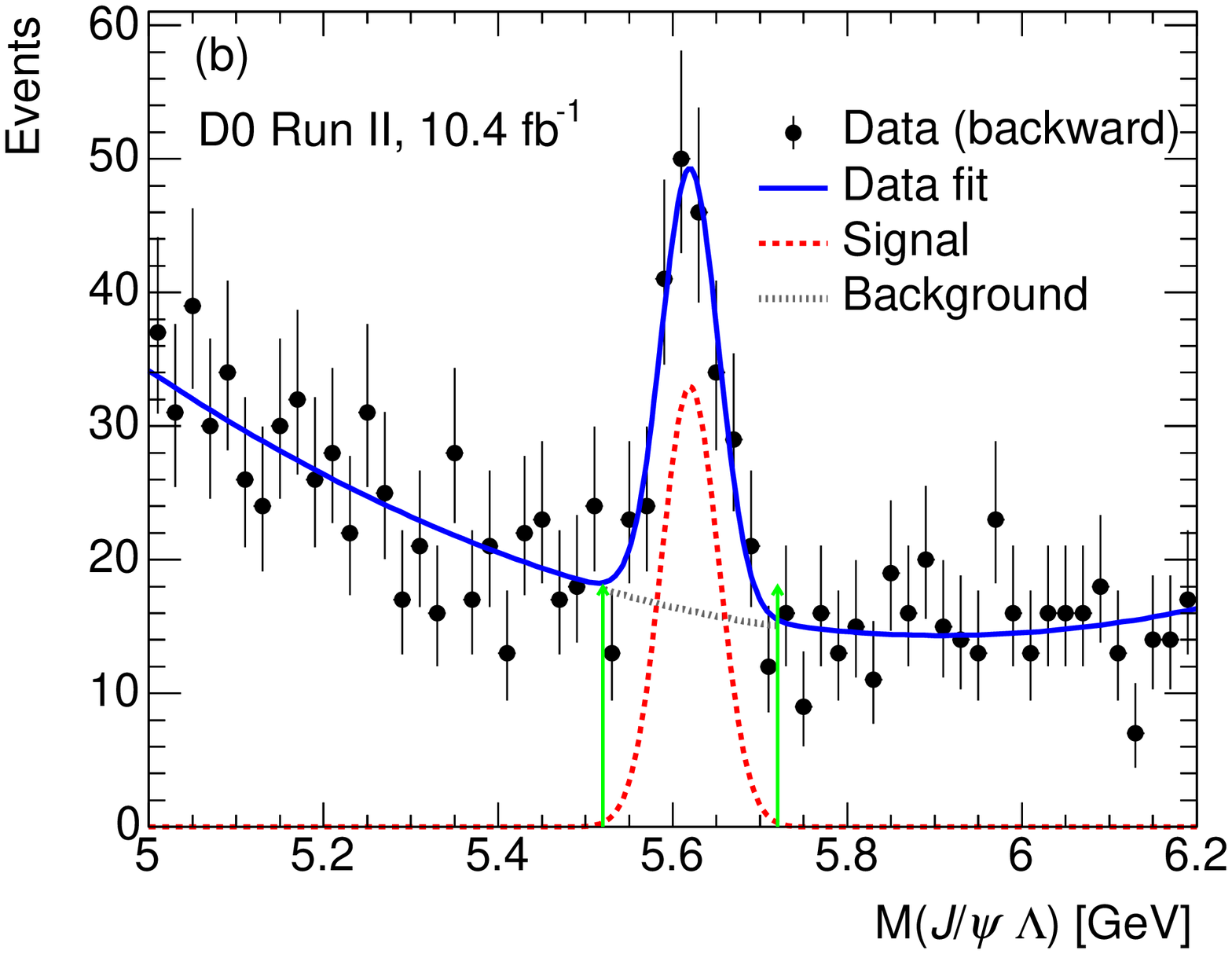}}
 \caption{(color online)  Invariant mass distribution of $\Lambda_b^0\rightarrow J/\psi \Lambda$ and $\overline  \Lambda_b^0\rightarrow J/\psi \overline \Lambda$  candidates in the rapidity range $0.5<|y|<1.0$ in the  (a) forward and (b) backward categories. The fit of a Gaussian signal function with a second-order Chebyshev polynomial background  function  is superimposed.  The vertical lines define the  signal region.
}
   \label{fig:mass}
\end{figure}

The  mass distributions for $\Lambda_b^0$ candidates in the  range $0.5<|y|<1.0$ in the forward and backward categories are shown in  Fig.~\ref{fig:mass}.  Binned maximum-likelihood fits  of a Gaussian signal  function  and a second-order Chebyshev polynomial for the  background yield a forward (backward) signal with a mean mass of $M(\Lambda_b^0)=5618.1 \pm 4.3$~MeV ($5619.9\pm4.7$~MeV), consistent with the world-average  $\Lambda_b^0$ mass~\cite{pdg2014}.
 The width depends on $y$ and varies between about $30$ and $50$~MeV. 
The average reconstructed $p_T$ of $\Lambda_b^0$ candidates is $\left< p_T\right> = 9.9$~GeV after background subtraction.

 The  production rates of  forward and backward $\Lambda_b^0$ and  $\overline \Lambda_b^0$ baryons are extracted from fits to the invariant mass distributions of forward and backward candidates in four rapidity bins in the range $0.1<|y|<2.0$, as defined in Table~\ref{tab:ratio}. We reject the  region $|y|<0.1$ where the asymmetry may be diluted by  forward-backward migration due to the finite  polar angle resolution~\cite{d0bplus}. 

 \begin{table*}[h!tb]
\caption{Efficiencies $\epsilon$, averaged values of background-subtracted transverse momenta $\left< p_T\right>$, backward and forward fitted yields for the signal $N(B)$ and  $N(F)$, forward-backward asymmetries $A$, and cross-section ratios $R$ in four intervals of rapidity.  Uncertainties on $\left< p_T\right>$, $N(B)$ and $N(F)$ are statistical only. 
Uncertainties on $\epsilon$  arise from the statistical precision of the simulated event samples.}
\begin{tabular}{c|c|c|c|c|c|c} \hline \hline
$|y|$ & $\epsilon$ (\%)  &  $\left< p_T\right>$ (GeV)  & $N(B)$        &  $N(F)$              & $A \pm {\rm  (stat)} \pm {\rm (syst)}$       &  $R \pm {\rm  (stat)} \pm {\rm (syst)}$  \\\hline
$0.1 - 0.5$ & $ 0.70 \pm 0.01$ & $10.2\pm0.1$     & $125\pm 18$   & $\hspace{5pt}92\pm 17$     &  $-0.15\pm0.11 \pm 0.03$  & $1.36 \pm 0.32 \pm 0.06$  \\    
$0.5 - 1.0$ & $1.01 \pm 0.01$  & $10.0\pm 0.1$    & $135\pm 19$   & $154\pm 22$     &  $\hspace{7.5pt}0.07\pm0.10 \pm 0.02$   & $0.88 \pm 0.18 \pm 0.04$  \\ 
$1.0 - 1.5$ & $0.97 \pm 0.01$  & $\hspace{5pt}9.7\pm 0.1$   & $123\pm 16$   & $158\pm 23$     &  $\hspace{7.5pt}0.12\pm0.10 \pm 0.02$    & $0.78 \pm 0.15 \pm 0.04$  \\   
$1.5 - 2.0$ & $0.32 \pm 0.01$  & $\hspace{5pt}9.8\pm 0.2$   & $22\pm 9$       & $\hspace{5pt}33\pm 10$     &  $\hspace{7.5pt}0.21\pm0.24 \pm 0.02$    & $0.67 \pm 0.34 \pm 0.03$   \\     \hline  \hline 
\end{tabular}
\label{tab:ratio}
\end{table*}

Samples of fully simulated Monte Carlo (MC) signal events are obtained at LO with {\sc pythia}~\cite{Pythia} and at NLO with {\sc MC@NLO}~\cite{mcatnlo}, using the parton distribution function sets {\sc CTEQ6L1} and {\sc CTEQ6M1}~\cite{CTEQ}, respectively. {\sc Pythia} generates $b\bar{b}$ quark pairs via direct $2 \rightarrow 2$ processes ($q_i \bar{q}_i,gg  \rightarrow b\bar{b}$) 
and  decays of gauge bosons, as well as through flavor excitation processes like $bg\rightarrow bg$, and gluon splittings, $g\rightarrow b\bar{b}$. The event generator   {\sc MC@NLO} is interfaced with {\sc Herwig}~\cite{Herwig} for parton showering and hadronization. After hadronization, bottom hadron decays are simulated with  {\sc EvtGen}~\cite{EvtGen}. In the simulation, the $\Lambda_b^0$ and $\overline{\Lambda}_b^0$ baryons are forced to  decay to $ J/\psi \Lambda$, $J/\psi \rightarrow \mu^+\mu^-$, using the phase space ({\sc PHSP}) and vector to lepton-lepton ({\sc VLL}) models in {\sc EvtGen}. The detector response is simulated with {\sc geant3}~\cite{Geant} and multiple $p\bar{p}$ interactions (pileup) are modeled by overlaying hits from random bunch crossings in data. 
A MC sample generated with  {\sc pythia}, 30 times the number of signal events in the data sample, is used to obtain efficiencies for reconstructing $\Lambda_b^0$ baryons in each of the four rapidity intervals shown in Table~\ref{tab:ratio}. 
The $\Lambda_b^0$ efficiencies are suppressed by the large transverse momentum requirement on the $\Lambda$ candidates and by the low reconstruction efficiency for the long-lived $\Lambda$ baryon.

 \begin{table}[h]
\begin{center}
\caption{Systematic uncertainties (in~\%) on the measurement of the backward-to-forward  ratio~$R$.}
\begin{tabular}{c|c}\hline \hline
Source     &   Uncertainty (\%)  \\ \hline
Signal shape        &      2               \\    
Background shape    &      2               \\     
Detection efficiency  &    3     \\ \hline 
Total syst. uncertainty               &      4     \\ \hline \hline
\end{tabular}
\label{tab:syst}
\end{center}
\end{table}

Most of the systematic uncertainties in the production cross-sections of $\Lambda_b^0$ and  $\overline \Lambda_b^0$ baryons  arise from uncertainties in the kinematic acceptance and detection efficiency of final-state particles and cancel  in the measurements of the  asymmetry $A$ and ratio $R$. The remaining uncertainties are due to the signal and background shapes assumed in the mass fits and  the different efficiencies of  forward and backward particle reconstruction. The uncertainty from the signal shape is estimated by comparing the results of the central fits with the results obtained when the width parameters for the forward and backward categories are constrained  to be equal.  The sensitivity to the background shape is estimated by increasing the lower mass requirement to $M(J/\psi \Lambda)>5.2$~GeV, thus excluding the mass range where feed-down from multi-body bottom baryon decays may be present. The estimate of the uncertainty on  the detection efficiency is based on the average deviation from unity of the ratio $R$ of reconstructed events in four rapidity intervals for a sample of MC events generated with no asymmetry.
Adding the  uncertainties in quadrature results in a total systematic uncertainty of  $\pm4$\%. 
The systematic uncertainties are summarized in Table~\ref{tab:syst}.

 The fitted signal yields and the resulting  forward-backward asymmetry $A$
  are presented in  Table~\ref{tab:ratio}. 
 We observe  that there is  a weak correlation between rapidity $y$ and the averaged value of background-subtracted transverse momentum $\left<p_T\right>$ of $\Lambda_b^0$ candidates.  
The asymmetry integrated over $|y|$, taking into account the rapidity dependent efficiency $\epsilon$,  is $A=0.04 \pm 0.07 {\rm \thinspace (stat)} \pm 0.02 {\rm \thinspace (syst)}$.

\begin{figure}[tb]
  \centering
  \includegraphics[width=0.9\columnwidth]{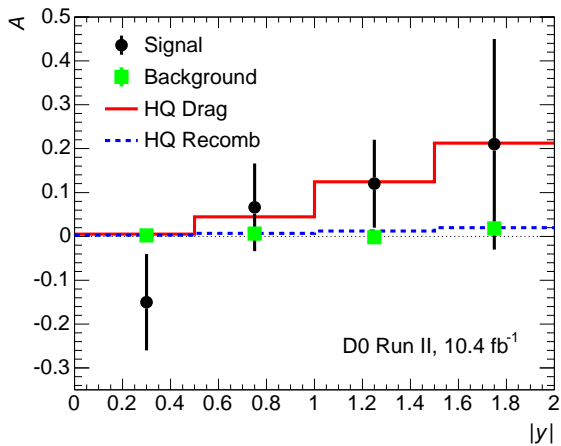}
  \caption{(color online) Measured forward-backward asymmetry $A$ versus rapidity $|y|$ compared to predictions of the heavy quark recombination
model~\cite{hqr} and a simulated effect of the longitudinal momentum shift due to beam drag (see Ref.~\cite{rosner} and text). The background asymmetry is obtained from $J/\psi \Lambda$ candidates in the $\Lambda_b^0$ mass sidebands (uncertainties are small compared to the symbol size).  Measurements are placed at the centers of the rapidity intervals defined in Table~\ref{tab:ratio}.
}
   \label{fig:ycomp}
\end{figure}

The  forward-backward asymmetry as a function  of $|y|$ is shown in  Fig.~\ref{fig:ycomp}. There is a wide range of model predictions for this asymmetry. The ``heavy quark recombination''  model~\cite{hqr}, as shown in  Fig.~\ref{fig:ycomp}, predicts a modest asymmetry, reaching
 $\approx2\%$ 
near $|y|=2$. While {\sc pythia} predicts no asymmetry, the {\sc MC@NLO} generator interfaced with {\sc Herwig}  predicts a large asymmetry, reaching 100\% close to $|y|=2$. Our results are consistent  with no asymmetry  within the large uncertainties, although they show a trend of increasing asymmetry with increasing  $|y|$ that could 
 be interpreted as the effect of the longitudinal momentum imparted to a  $\Lambda_b^0$ or  $\overline \Lambda_b^0$ particle by the beam remnant.
Assuming a shift  of  $\Delta p_z = 1.4$~GeV in the particle longitudinal momentum, as estimated by Rosner~\cite{rosner},
  we have simulated the effect by adding $1.4$ GeV  ($-1.4$ GeV)  to the $\Lambda_b^0$   ($\overline \Lambda_b^0$)   baryon $p_z$ in the generated {\sc pythia}  events.
As shown in  Fig.~\ref{fig:ycomp}, our result is in a good  agreement with  this prediction.
 We find our results in disagreement with the large asymmetry predicted by {\sc MC@NLO}+{\sc Herwig}.

\begin{figure}[tb]
  \centering 
 \includegraphics[width=0.9\columnwidth]{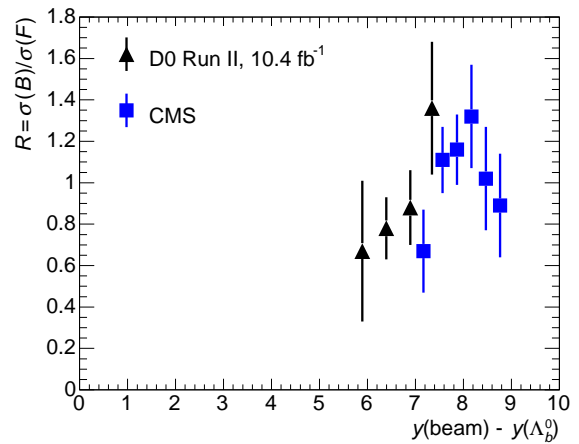}
  \caption{Measured ratio of the backward to forward production cross sections  versus rapidity loss
 compared to the  $\overline \Lambda_b^0$ to $ \Lambda_b^0$ production cross section ratio at CMS
taken from Table II of Ref.~\cite{cms}.
Measurements are placed at the centers of their rapidity loss ranges. 
}
   \label{fig:rcompd0cms}
\end{figure}

The results for the backward-to-forward ratio $R$ for the same rapidity intervals are given in Table~\ref{tab:ratio} and shown in Fig.~\ref{fig:rcompd0cms}, where  
we compare with the results for the ratio of cross sections, $\sigma(\overline \Lambda_b^0)/\sigma(\Lambda_b^0)$, for the six rapidity bins reported by the CMS Collaboration~\cite{cms}.
All results are presented as functions of the ``rapidity loss'', defined as the difference between the rapidity of the beam, $y(\text{beam}) = 7.64$~(8.92) at the Tevatron (LHC),  and the rapidity $y$ of the $\Lambda_b^0$ baryon. 
   The D0 and CMS   results are consistent within large uncertainties. 
 Together, they show a trend of $R$ to fall with increasing rapidity and decreasing rapidity loss. 
 The D0 result for the ratio $R$ integrated over rapidity, taking into account the rapidity dependent efficiency $\epsilon$, is
 $R=0.92 \pm 0.12 {\rm \thinspace (stat)} \pm 0.04 {\rm \thinspace (syst)}$, to be compared with the value of $R=1.02 \pm 0.07 {\rm \thinspace (stat)} \pm 0.09 {\rm \thinspace (syst)}$ reported by the CMS Collaboration.

In order to verify that detector effects on $R$ and $A$  are not significant, the analysis was repeated considering
candidates with $y > 0$   (or $y < 0$) only, and 
 $\Lambda_b^0$ (or  $\overline \Lambda_b^0$) only.  
Within statistical uncertainties, all results are consistent with each other and with the measurements listed in Table~\ref{tab:ratio}. 
Furthermore, as shown in Fig.~\ref{fig:ycomp}, we find a negligible forward-backward asymmetry in the four intervals of rapidity in a sample of background candidates obtained from the $\Lambda_b^0$ mass sidebands (region above and below the  $\Lambda_b^0$ signal region defined in Fig.~\ref{fig:mass}) with no $L_{xy}$ requirement.


In summary, we have presented a measurement of the forward-backward asymmetry in  the production of $\Lambda_b^0$ and $\overline \Lambda_b^0$ baryons as a function of rapidity $|y|$. Together with related results from the LHC, the data show a tendency of forward particles that share valence quarks with beam remnants, to be emitted at  larger values of rapidity, corresponding to smaller rapidity loss, than their backward counterparts. 
The measured ratio of the backward-to-forward production rate  at the mean transverse momentum of $\left< p_T\right> = 9.9$~GeV, averaged
over rapidity in the range $0.1<|y|<2.0$,  is  $R=0.92\pm 0.12 {\rm \thinspace (stat)} \pm 0.04 {\rm \thinspace (syst)}$. The measured forward-backward asymmetry  is $A=0.04 \pm 0.07 {\rm \thinspace (stat)} \pm 0.02 {\rm \thinspace (syst)}$.
  

%

We would like to thank W.~K.~Lai and A.~K.~Leibovich for providing predictions of the heavy quark recombination model for the D0 kinematic range, and J.~L. Rosner for useful discussions.
We thank the staffs at Fermilab and collaborating institutions
and acknowledge support from the
Department of Energy and National Science Foundation (United States of America);
Alternative Energies and Atomic Energy Commission and
National Center for Scientific Research/National Institute of Nuclear and Particle Physics  (France);
Ministry of Education and Science of the Russian Federation, 
National Research Center ``Kurchatov Institute" of the Russian Federation, and 
Russian Foundation for Basic Research  (Russia);
National Council for the Development of Science and Technology and
Carlos Chagas Filho Foundation for the Support of Research in the State of Rio de Janeiro (Brazil);
Department of Atomic Energy and Department of Science and Technology (India);
Administrative Department of Science, Technology and Innovation (Colombia);
National Council of Science and Technology (Mexico);
National Research Foundation of Korea (Korea);
Foundation for Fundamental Research on Matter (The Netherlands);
Science and Technology Facilities Council and The Royal Society (United Kingdom);
Ministry of Education, Youth and Sports (Czech Republic);
Bundesministerium f\"{u}r Bildung und Forschung (Federal Ministry of Education and Research) and 
Deutsche Forschungsgemeinschaft (German Research Foundation) (Germany);
Science Foundation Ireland (Ireland);
Swedish Research Council (Sweden);
China Academy of Sciences and National Natural Science Foundation of China (China);
and
Ministry of Education and Science of Ukraine (Ukraine).
%


\end{document}